$Ni_2Mo_3O_8$: zig-zag antiferromagnetic order an integer spin non-centrosymmetric honeycomb lattice

Jennifer R. Morey,[a,b] Allen Scheie,[b] John P. Sheckelton,[a,b] Craig M. Brown,[c] and Tyrel M. McQueen[a,b,d]

[a] Department of Chemistry, Johns Hopkins University, Baltimore, Maryland 21218, USA

[b] Institute for Quantum Matter, Department of Physics and Astronomy, Johns Hopkins University, Baltimore, Maryland 21218, USA

[c] National Institute for Standards and Technology, Gaithersburg, MD 20899 USA and Department of Chemical and Biomolecular Engineering, University of Delaware, Newark, Delaware, 19716 USA

[d] Department of Materials Science and Engineering, Johns Hopkins University, Baltimore, Maryland 21218, USA

Abstract

Theoretical studies have predicted the existence of topological magnons in honeycomb compounds with zig-zag antiferromagnetic (AFM) order. Here we report the discovery of zig-zag AFM order in the layered and non-centrosymmetric honeycomb nickelate $Ni_2Mo_3O_8$ through a combination of magnetization, specific heat, x-ray and neutron diffraction and electron paramagnetic resonance measurements. It is the first example of such order in an integer-spin non-centrosymmetric structure ($P6_3mc$). Further, each of the two distinct sites of the bipartite honeycomb lattice has a unique crystal field environment, octahedral and tetrahedral $Ni^{2+}$ respectively, enabling independent substitution on each sublattice. Replacement of Ni by Mg on the octahedral site suppresses the long range magnetic order and results in a weakly ferromagnetic state. Conversely, substitution of Fe for Ni enhances the AFM ordering temperature. Thus $Ni_2Mo_3O_8$ provides a platform on which to explore the rich physics of $S = 1$ on the honeycomb in the presence of competing magnetic interactions with a non-centrosymmetric, formally piezeo-polar, crystal structure.

A.    Introduction

The prediction and discovery of topological phenomena in materials has ignited a global search for new quantum materials and states of matter [1, 2], with potential applications in quantum computing and information storage. The physical realization of theoretically proposed topological states requires the ability to produce materials with highly controlled structural, electronic, and magnetic properties. Most materials release inherent magnetic degeneracy at sufficiently low temperatures by mechanisms such as structural phase transitions, local magnetic ordering, and changes in the degree of electron localization (e.g. by formation of singlet pairs with neighboring ions), but there are some states of matter postulated to retain finite degeneracy to $T = 0$ K, such as quantum spin liquids (QSLs) [3-6].

One of the main structure types known to host quantum frustrated magnetic topological phenomena is the 'honeycomb' structure, which is a two dimensional bipartite lattice. Unlike the triangular lattice or spinel structure, the honeycomb is not inherently geometrically frustrated, but becomes frustrated in the presence of longer range magnetic interactions or anisotropic magnetic exchanges.

One example of this is the ruthenium honeycomb in α-$RuCl_3$ which may host almost exactly the interactions that would allow a finite degenerate quantum spin liquid (Kitaev QSL) state to emerge [7-11]; It is suggested that it is the strong next-nearest neighbor and next-next-nearest neighbor interactions that stabilize the frustration [12, 13]. Furthermore, extensive experimental and theoretical investigations into iridium honeycomb compounds $Li_2IrO_3$ [14-18] and $Na_2IrO_3$ [19-23] have realized many of the types of magnetically ordered states that are proximal to QSL states – i.e. stripy antiferromagnetic (AFM), zig-zag AFM, and Néel AFM [24-29].

The nature of the spin interaction, relevant magnetic exchanges, structural geometry, order, symmetry, and spin orbit coupling (SOC) influence the magnetic ground state of a compound. SOC generally increases with atomic number and becomes a controlling factor in 4d and 5d transition metal honeycombs, particularly those incorporating iridium and ruthenium. Strong SOC has been posited as the reason that iridium honeycombs have a ground state that is magnetically ordered rather than a QSL [29].

Despite having weaker SOC than the 4d or 5d equivalents, 3d ions with strong anisotropy, e.g. $Co^{2+}$, may also harbor strong bond-dependent interactions between ions [30, 31]. Further, recent theoretical predictions have shown that honeycomb compounds with zig-zag AFM and stripy AFM order may host topologically non-trivial magnons that are robust under Dzyaloshinskii-Moriya interactions [32, 33].

Here we report that $Ni_2Mo_3O_8$, which contains a honeycomb of $S = 1$ $Ni^{2+}$ ions and has previously been reported to remain paramagnetic down to $T = 2$ K [34], undergoes a transition to a zig-zag ordered antiferromagnetic state below $T_N = 6$ K, and is thus a candidate for harboring topological excitations. Compared to other nickel compounds known to have zig-zag antiferromagnetic order, including $BaNi_2V_2O_8$, $BaNi_2As_2O_8$, $Na_3Ni_2BiO_6$, $A_3Ni_2SbO_6$ (A = Li, Na), and $Cu_3Ni_2SbO_6$ [35-37], $Ni_2Mo_3O_8$ is unique in that the two triangular sublattices of the honeycomb have different local coordination environments of the $Ni^{2+}$ ions (octahedral and tetrahedral), permitting selective replacement of one of the two halves of the bipartite lattice. Additionally, it is the first example of zig-zag AFM order in a non-centrosymmetric $S = 1$ honeycomb material, complementing the only other known non-centrosymmetric zig-zag antiferromagnetic material, $Na_2Co_2TeO_6$, with $S = 3/2$.

In $Ni_2Mo_3O_8$, we find substitution of non-magnetic $Mg^{2+}$ on the tetrahedral site removes long range magnetic order, with remnant small ferromagnetic interactions between $Ni^{2+}$ ions. In contrast, substitution of $S = 2$ $Fe^{2+}$ for $Ni^{2+}$ results in a large increase in the antiferromagnetic ordering temperature to $T_N = 50$ K. $Ni_2Mo_3O_8$ is a realization of zig-zag order in a non-centrosymmetric antiferromagnet; The ability to selectively substitute one of the two sites in the honeycomb make this material an excellent platform to use to investigate the underlying physics of the selection of magnetic ground states on the honeycomb lattice.

B.  Experimental Methods

*1.  Powder Synthesis*

$M_2Mo_3O_8$, $M$ = (Mg, Ni, Fe, Zn), were synthesized by intimately mixing $M$O or $M_2O_3$ and $MoO_2$ with a small stoichiometric excess of $M$O where $M$ = (Mg, Ni) in an agate mortar and pestle, followed by compression into a pressed pellet and sealing in an alumina crucible in a quartz ampoule evacuated to $10^{-2}$ mmHg. The samples were first heated at 200 °C/hr to 950 °C, held at that temperature overnight, and then quenched by removal of the quartz ampoule from the furnace to the benchtop to cool. Successive regrinding, repressing, resealing, and overnight reheating cycles, with the sample placed directly into and removed from a furnace at $T$ = 950 °C, were performed until phase purity was achieved. Purity was checked with Rietveld refinements of powder X-ray diffraction (PXRD) patterns.

*2.  Nuclear and Magnetic Structural Characterization*

PXRD patterns were collected on a Bruker D8 Focus diffractometer with a LynxEye detector using Cu Kα radiation. Rietveld refinements were performed using Topas 4.2 (Bruker). Neutron powder diffraction (NPD) experiments on $Ni_2Mo_3O_8$ and $MgNiMo_3O_8$ were performed at the National Institute for Standards and Technology Center for Neutron Research (NCNR) on the BT-1 powder diffractometer using the Ge311 monochromator, 60' collimation, and a wavelength $\lambda_{neutron}$ = 2.0775 Å. Nuclear structural refinements were performed using GSAS [38] and EXPGUI [39] and cross referenced with structural refinements done in the FullProf Suite [40]. Time of flight neutron powder diffraction experiments were done at the high resolution powder diffractometer POWGEN at Oak Ridge National Laboratory using Frame 1.5 at $T$ = 10 K and $T$ = 300 K. LeBail unit cell refinements were used to account for starting material (NiO, MgO,

MoO$_2$) and side product (NiMoO$_4$) impurities, present at the < 2 % level.

The magnetic unit cell was manually indexed using GSAS and EXPGUI and confirmed using k-search in the FullProf suite. SARAh Representational Analysis software [41] and FullProf were used in tandem to determine the final structure. Structures were visualized using Vesta software [42].

### 3.   Physical Properties Characterization

Magnetization and heat capacity measurements were done using a Quantum Design Physical Properties Measurement System. Temperature dependent magnetization data were collected from $T = (2 - 300)$ K under applied fields of $\mu_0 H = 0.5$ T and 1 T. Susceptibility was computed as $\chi = \Delta M / \Delta H$ numerically from the two fields for each temperature. The 0.5 T and 1 T fields were chosen as representative of a linear portion of the magnetization curve. Curie-Weiss analysis was performed over the temperature range 150 K < $T$ < 300 K after linearization of susceptibility data with a temperature independent $\chi_0$.

Zero field heat capacity was collected from $T = 2$ K to $T = 300$ K for Ni$_2$Mo$_3$O$_8$ and to $T = 150$ K for MgNiMo$_3$O$_8$ and FeNiMo$_3$O$_8$ using the semi-adiabatic pulse technique with a 2 % temperature rise and measurement over 3 time constants in time.  Measurements were performed in triplicate. Field-dependent heat capacity was collected up to $\mu_0 H = 5$ T from $T = 2$ K to $T = 20$ K. Ni$_2$Mo$_3$O$_8$ and MgNiMo$_3$O$_8$ were measured as pressed pellets, while FeNiMo$_3$O$_8$ was pressed with clean silver powder. Heat capacity measurements in the $T = 150$ mK – 3.5 K range were done on a Quantum Design Dilution Refrigerator (DR) using the semi-adiabatic pulse technique with a 2 % temperature rise and measurement over 3 time constants in time. Measurements were performed in triplicate. DR samples were pressed with clean silver powder to enhance thermal conductivity with the stage. In both cases, the heat capacity of silver was measured and subtracted from the raw signal.

The phononic contribution of $Ni_2Mo_3O_8$ was found by scaling the measured heat capacity of $Zn_2Mo_3O_8$ for the mass difference between nickel and zinc [43]. Similarly, the phononic contribution to the heat capacity of $MgNiMo_3O_8$ was found as the average of measurements on $Mg_2Mo_3O_8$ and $Zn_2Mo_3O_8$, scaled to account for the mass differences in the stoichiometric formulae. Literature reports on $Fe_2Mo_3O_8$ were used to scale measurements taken on $Zn_2Mo_3O_8$ manually to find the phonon contribution in $FeNiMo_3O_8$ [44].

### 4. Calculation Methods

The energy splitting of the $Ni^{2+}$ ions was calculated with a point charge model [45] using the PyCrystalField software package [46]. We built crystal electric field models using the ligand positions determined from the neutron diffraction experiments, and calculated the eigenstates of a single-ion Hamiltonian with crystal fields and spin orbit coupling treated non-perturbatively. More details are given in the Supplementary Information (SI).

### C. Results

### 5. Nuclear Structural Determination

$Ni_2Mo_3O_8$, $MgNiMo_3O_8$, and $FeNiMo_3O_8$ are isostructural: alternating layers of hexagonal honeycomb and trimerized molybdenum oxide layers. Analyses of NPD (Fig. 1(a-b)) and PXRD patterns support that $Ni_2Mo_3O_8$, $MgNiMo_3O_8$, and $FeNiMo_3O_8$ crystallize in the non-centrosymmetric hexagonal space group 186, $P6_3mc$, Table I.

The honeycomb lattice is a bipartite lattice comprised of two triangular sublattices. In $Ni_2Mo_3O_8$, one triangular sublattice is octahedrally coordinated $Ni^{2+}$ and the other is tetrahedrally coordinated $Ni^{2+}$, making this material an integer-spin honeycomb (Fig. 1c)). In $MgNiMo_3O_8$, 86(3) % of the *2b* octahedral sites and 14(3) % of the *2b* tetrahedral sites are occupied by nickel, and 14(3) % and 86(3) % of these sites, respectively, are occupied by non-magnetic magnesium ions. The sensitivity of the fit statistics to changes in stoichiometry is discussed in the SI. At $T = 15$ K, the oxygen ligands on the *2b* Wycoff position in $Ni_2Mo_3O_8$ are slightly distorted in the *c*-direction from their ideal positions around the nickel sites. In the octahedron, the O-Ni-O angle is 88.2(2) ° rather than the ideal 90°. In the tetrahedron, the O-Ni-O angle is 114.52(14) °, rather than the ideal 109.5°. This distortion has an anisotropic temperature dependence, shown in Fig. 2. The *c* lattice parameter decreases almost linearly from $T = 300$ K to $T = 15$ K, while the *a* lattice parameter decreases more rapidly than *c* from $T = 300$ K to $T \sim 180$ K and remains relatively constant from $T = 150$ K to $T = 15$ K. The ratio of the lattice parameters *a/c* over temperature in the lower panel of Fig. 2 is particularly instructive: it increases from $T = 300$ K to $T \sim 180$ K and decreases from $T = 130$ K to $T = 15$ K.

The oxygen ligand crystal field environment is similarly distorted in $MgNiMo_3O_8$ as it is in $Ni_2Mo_3O_8$. In these materials, the oxygen locations can be precisely located due to the scattering factor contrast available by NPD measurements. $FeNiMo_3O_8$ was characterized using PXRD; the best refinements are obtained with the octahedral site selectively occupied by $Fe^{2+}$ (Table I, refinements are plotted in Fig. 6, SI). The placement of $Fe^{2+}$ on the octahedral site somewhat surprising: while the ionic radius of $Ni^{2+}$ is slightly smaller than that of $Fe^{2+}$ (high spin) in both CN = 4, respectively 0.55 pm and 0.63 pm, and CN = 6, 0.69 pm and 0.79 pm, which would tend to favor placement of $Fe^{2+}$ on the octahedral site, crystal field stabilization energies would favor $Ni^{2+}$ on the octahedral site. Despite this expectation, other data is also consistent with an ordering of the $Fe^{2+}$ and $Ni^{2+}$ ions: there is a sharp antiferromagnetic transition in the susceptibility (see below), which would not be expected if $Fe^{2+}$ and $Ni^{2+}$ were randomly mixed. Thus we assume ordering of $Fe^{2+}$ and $Ni^{2+}$, but note that site mixing at the 10 % to 20 % level cannot be ruled out.

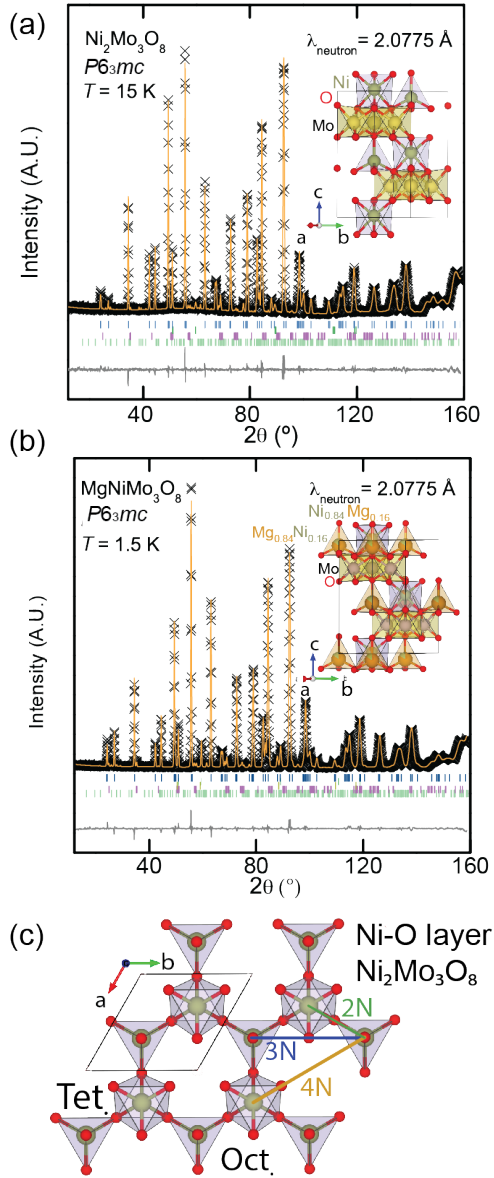

FIG. 1 Neutron powder diffraction patterns of (a) $Ni_2Mo_3O_8$ and (b) $MgNiMo_3O_8$, refined to the *P6_3mc* space group; Table I. Tick marks in descending vertical display order: $Ni_2Mo_3O_8$ (dark blue), NiO (dark green); MgO (brown); $MoO_2$ (purple), and $NiMoO_4$ (light green). MgO is not present in the refinement for $Ni_2Mo_3O_8$. (c) Top-down view of the nickel honeycomb lattice, showing alternating adjacent octahedrally and tetrahedrally coordinated atoms and nearest neighbor (2N; 3.384(3) Å), next nearest neighbor (3N; 5.759(5) Å) interactions, and next-next nearest neighbor (4N; 6.680(5) Å) interactions. Values in parentheses indicate one standard deviation in the final figures.

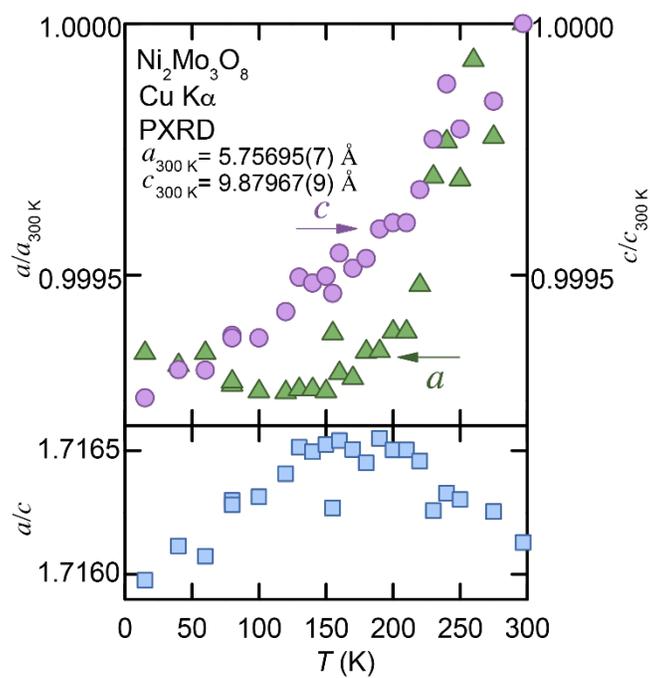

FIG. 2 Top panel: temperature dependence of the $a$ and $c$ lattice parameters of $Ni_2Mo_3O_8$ relative to $T = 300$ K values of 5.75695(7) Å and 9.87967(9) Å, respectively. Bottom panel: temperature dependence of the ratio of the lattice parameters.

TABLE I Atomic parameters for structural refinement of $(M1)(M2)Mo_3O_8$, $M1$ = (Ni, Mg, Fe), $M2$ = Ni; $Ni_2Mo_3O_8$ and $MgNiMo_3O_8$ from NPD (BT-1) at $T$ = 1.5 K and $T$ = 15 K respectively with $\lambda_{neutron} = 2.0775$ Å, $FeNiMo_3O_8$ from PXRD at room temperature with $\lambda_{Cu,K\alpha} = 1.5406$ Å. Occupancies of M1 and M2 are given as (Mg or Fe)/Ni and Ni/(Mg or Fe) respectively. Values in parentheses indicate one standard deviation in the final figures.

|  |  | $Ni_2Mo_3O_8$ | $MgNiMo_3O_8$ | $FeNiMo_3O_8$ |
|---|---|---|---|---|
|  | $a$ (Å$^2$) | 5.74683(5) | 5.75166(3) | 5.76580(2) |
|  | $c$ (Å$^2$) | 9.8626(2) | 9.85620(9) | 9.90929(3) |
|  | $T$ (K) | 15 | 1.5 | 295 |
| M1 | $x$ | 1/3 | 1/3 | 1/3 |
| 2b | $y$ | 2/3 | 2/3 | 2/3 |
|  | $z$ | 0.9480(4) | 0.9452(2) | 0.9715(2) |
|  | $U_{iso}$ | 0.0057(7) | 0.0006(4) | 0.0109(3) |
|  | Occ. | 1 | 0.86/0.14(3) | 1.0(1)/0.0 |
| M2 | $x$ | 1/3 | 1/3 | 1/3 |
| 2b | $y$ | 2/3 | 2/3 | 2/3 |
|  | $z$ | 0.5116(3) | 0.5120(5) | 0.5348(2) |
|  | $U_{iso}$ | 0.0056(8) | 0.00106(4) | 0.0109(3) |
|  | Occ. | 1 | 0.86/0.14(3) | 1.0(1)/0 |
| Mo | $x$ | 0.1440(2) | 0.14586(9) | 0.14688(3) |
| 6c | $y$ | -0.1440(2) | -0.14586(9) | -0.14688(3) |
|  | $z$ | 0.2489(2) | 0.25017(14) | 0.2733(10) |
|  | $U_{iso}$ | 0.0042(7) | 0.0002(2) | 0.0058(2) |
| O1 | $x$ | 0 | 0 | 0 |
| 2a | $y$ | 0 | 0 | 0 |
|  | $z$ | 0.6839(5) | 0.3890(3) | 0.6165(4) |
|  | $U_{iso}$ | 0.008(2) | 0.0095(8) | 1 |
| O2 | $x$ | 1/3 | 1/3 | 1/3 |
| 2b | $y$ | 2/3 | 2/3 | 2/3 |
|  | $z$ | 0.1461(4) | 0.147(2) | 0.1765(4) |
|  | $U_{iso}$ | 0.0012(13) | 0.0003(5) | 1 |
| O3 | $x$ | 0.4880(3) | 0.4878(2) | 0.4882(2) |
| 6c | $y$ | -0.4880(3) | -0.4878(2) | -0.4882(2) |
|  | $z$ | 0.3659(3) | 0.36774(17) | 0.3971(4) |
|  | $U_{iso}$ | 0.0044(4) | 0.0047(3) | 1 |
| O4 | $x$ | 0.1688(3) | 0.1723(2) | 0.1665(3) |
| 6c | $y$ | -0.1688(3) | -0.1723(2) | -0.1665(3) |
|  | $z$ | 0.6342(3) | 0.36774(17) | 0.6609(2) |
|  | $U_{iso}$ | 0.0015(7) | 0.0173(4) | 1 |
|  | $wRp$ | 0.0715 | 0.0415 | 2.88 |
|  | $Rp$ | 0.0521 | 0.0288 | 2.23 |
|  | $\chi^2$ | 2.526 | 3.913 | 1.41 |

6. *Physical Properties*

$Ni_2Mo_3O_8$ and $MgNiMo_3O_8$ both exhibit a peak in heat capacity at $T \approx 6$ K, Fig.3 (a,b). It is at slightly higher temperature and is sharper in $Ni_2Mo_3O_8$, which is consistent with this material being less disordered and having stronger magnetic interactions than $MgNiMo_3O_8$. The application of a $\mu_0H = 5$ T magnetic field causes the peak to shift to lower temperatures in $Ni_2Mo_3O_8$ and to higher temperatures in $MgNiMo_3O_8$, which is indicative of antiferromagnetic and ferro/ferrimagnetic orders, respectively.

Strikingly, $Ni_2Mo_3O_8$ and $MgNiMo_3O_8$ recover the same amount of entropy per magnetic ion by $T \sim 150$ K. The entropy loss looks to be two step: one degree of freedom is lost between $T = 10$ K and $T = 150$ K and two more at the $T \sim 6$ K transition. The high temperature phonon contribution, calculated from the mass-adjusted measured heat capacity of non-magnetic analogs, describes the high temperature behavior of the materials well. This is highlighted in the insets, which are plotted on a linear temperature scale. There is a large peak in the heat capacity of $FeNiMo_3O_8$ at $T \sim 50$ K that recovers $\Delta S = 20.54(5)$ J mol$^{-1}$ K$^{-1}$, between $T = 2$ K and $T = 100$ K, Fig. 4. The phononic background is consistent with reports on the related compound $Fe_2Mo_3O_8$ [44]. The changes in entropy of all three compounds are summarized in Table II.

TABLE II Summary of recovered entropy per formula unit (f.u.), shown in Fig. 3(c) and the lower panel of Fig.4.

|  | $\Delta S_{mag}$ (J mol-f.u$^{-1}$.K$^{-1}$) |
|---|---|
| $Ni_2Mo_3O_8$ | 13.9(7) |
| $MgNiMo_3O_8$ | 6.9(3) |
| $FeNiMo_3O_8$ | 20.5(1.0) |

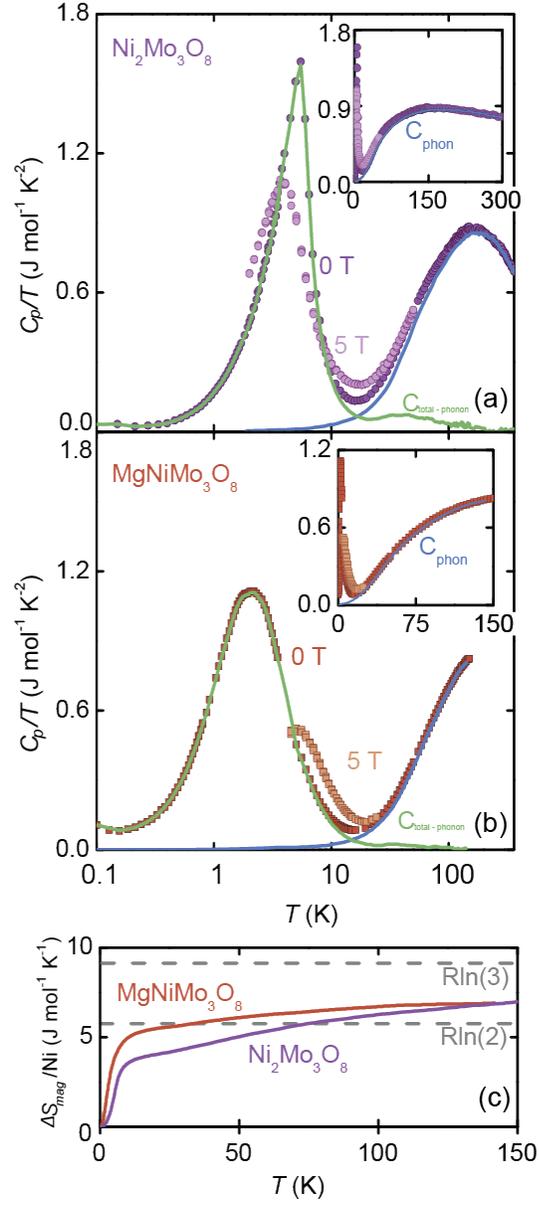

FIG. 3 (a) Heat capacity over temperature versus the logarithm of temperature of $Ni_2Mo_3O_8$ (top panel, purple circles) and (c) $MgNiMo_3O_8$ (brown squares). Magnetic heat capacity (green curve) calculated by subtracting the phononic contribution (blue curve) calculated from measured non-magnetic analog materials. Insets: Heat capacity over temperature versus linear temperature, highlighting the high temperature phonon contribution. (c) Entropy as a function of temperature.

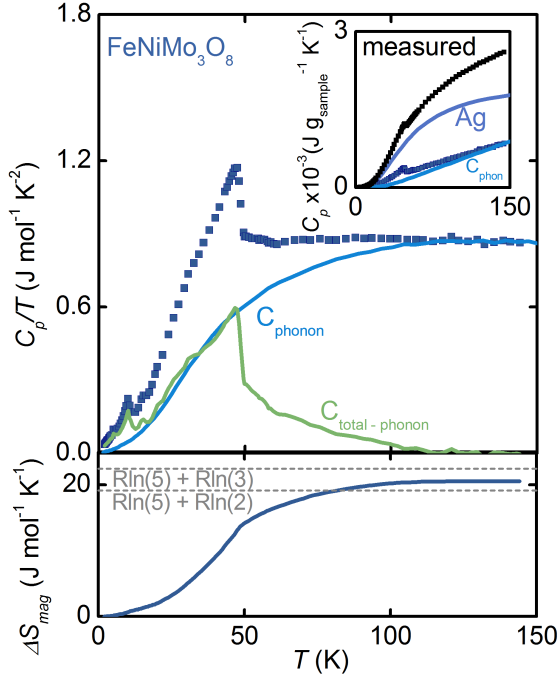

FIG. 4 Top: Heat capacity over temperature versus temperature of $FeNiMo_3O_8$ measured from $T = 2$ to $T = 150$ K (dark blue squares). Inset: Raw measured data (black squares) included heat capacity from clean silver powder pressed with the sample (blue curve), which was subtracted to isolate only the contribution from $FeNiMo_3O_8$. A peak at 50 K capturing between $R\ln(5) + R\ln(2)$ and $R\ln(5) + R\ln(3)$ of entropy (bottom panel, dark blue curve) was determined to be magnetic (green curve, top panel) by subtracting the phonon contribution to the specific heat (light blue curve, top panel and inset, from measured non-magnetic analog $Zn_2Mo_3O_8$, scaled to be consistent with literature measurements on $Fe_2Mo_3O_8$[44]).

All three compounds exhibit Curie-Weiss behavior at $T > 100$ K, Fig.5(a). $MgNiMo_3O_8$ has a small positive Weiss temperature of $\theta_W = 6.5(1.3)$ K, consistent with weak ferromagnetic interactions, and a Curie constant of 1.280(7) and $p_{eff} = 3.20(3)$ $\mu_B$. $Ni_2Mo_3O_8$ has a larger negative Weiss temperature of $\theta_W = -55.5(5)$ K, consistent with antiferromagnetic interactions, a total Curie constant of 5.518(1.0), and an average $p_{eff}$ of 4.70(3) $\mu_B$ per nickel atom, summarized in Table III. $FeNiMo_3O_8$ exhibits a clear antiferromagnetic phase transition at $T \sim 50$ K, Fig.

5(b). The effective magnetic moment is 6.86(4) $\mu_B$, which is close to the expected spin-only moment of 7.32 $\mu_B$ of combined high-spin $Fe^{2+}$ (4.49 $\mu_B$) and $Ni^{2+}$ (2.83 $\mu_B$). The Weiss temperature is $T$ = -101.5(3) K, indicating strong antiferromagnetic interactions.

At $T$ = 2 and $T$ = 6 K, the field dependent magnetization of $Ni_2Mo_3O_8$ has a metamagnetic curvature which is not visible at $T$ = 15 K, Fig. 5 (a) inset. Such metamagnetism suggests a low-lying (in field) magnetic phase transition is possible. This behavior could be interpreted as differences in in-plane and out-of-plane magnetic responses, for which single crystal samples are necessary to fully understand the nature of the transition [8]. There is no apparent hysteresis to the curve, suggesting that there is little to no ferromagnetic component of the magnetization at this temperature. The magnetic response of $MgNiMo_3O_8$ fits well to a Brillouin function in the $T$ = 2 K to $T$ = 300 K temperature range and is thus likely paramagnetic at all measured temperatures (Fig. 3 and Table II, SI).

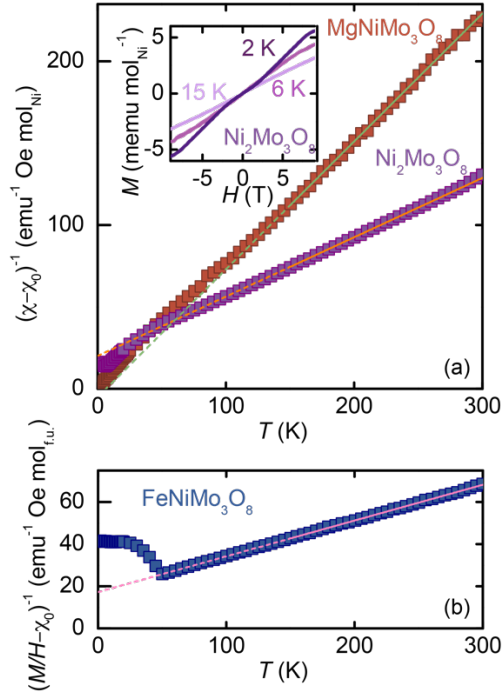

FIG. 5 Inverse susceptibility of $Ni_2Mo_3O_8$, $MgNiMo_3O_8$, and $FeNiMo_3O_8$ linearized and fit to the Curie-Weiss law in the temperature range of $T = 150$ K to 300 K, fit values summarized in Table III. (a) Inverse susceptibility of $MgNiMo_3O_8$ (brown squares) is non-linear below $T = 150$ K but shows no clear ordering transition. In contrast, a small upturn at $T = 6$ K in the inverse susceptibility of $Ni_2Mo_3O_8$ (purple squares) indicates an antiferromagnetic phase transition. The inverse susceptibility of this material is also non-linear in the $T = 6$ K to 150 K temperature range. Inset: Magnetization versus applied field of $Ni_2Mo_3O_8$ at $T = 2$ K, 6 K, and 15 K. (b) A sharp uptick in the inverse susceptibility of $FeNiMo_3O_8$ indicates a clear antiferromagnetic phase transition at $T \sim 50$ K. 1 Oe = $(1000/4\pi)$ A/m and 1 emu/(mol Oe) = $4\pi \ 10^{-6}$ m$^3$/mol.

TABLE III Fit values for Curie-Weiss analysis of high temperature magnetic susceptibility of $Ni_2Mo_3O_8$, $MgNiMo_3O_8$ and $FeNiMo_3O_8$, shown graphically in Fig. 5. $C$ and $p_{eff}$ are per formula unit. 1 Oe = $(1000/4\pi)$ A/m.

|  | $Ni_2Mo_3O_8$ | $MgNiMo_3O_8$ | $FeNiMo_3O_8$ |
|---|---|---|---|
| $C$ (emu K mol$^{-1}$ K$^{-1}$) | 5.52(1.4) | 1.28(7) | 5.89(9) |
| $p_{eff}$ ($\mu_B$) | 6.64(6) | 3.20(3) | 6.86(4) |
| $\theta_W$ (K) | -55.5(5) | 6.5(1.3) | -101(1.0) |
| $T_N$ (K) | 6.0(2) | - | 50.0(2) |
| $\chi_0$ (emu mol$^{-1}$ Oe$^{-1}$) | 0.0025 | 0.0015 | 0.00055 |

7.  *Electron Spin Resonance*

The ESR data in Fig. 7 (a) and (b) from $Ni_2Mo_3O_8$ and $MgNiMo_3O_8$ have broad resonances, which is typical of $S = 1$ systems [47]. There are two magnetic sites in each unit cell: the octahedrally coordinated and tetrahedrally coordinated nickels on the two triangular honeycomb sublattices. In $Ni_2Mo_3O_8$, these sites are equally populated. In $MgNiMo_3O_8$, 14 % of the tetrahedral sites and 86 % of the octahedral sites are populated by Ni (determined from NPD), and the remaining sites are non-magnetic. Thus, the ESR data from $Ni_2Mo_3O_8$ should show two equally-weighted resonances and the data from $MgNiMo_3O_8$ should show two resonances at 14 % and 86 % on each of the respective sites. This is visually consistent with the data, shown in Fig. 6, $Ni_2Mo_3O_8$, and Fig. 7, $MgNiMo_3O_8$. The resonance for $Ni_2Mo_3O_8$ looks like one broad resonance, which can be decomposed into two similarly-sized overlapping features. The resonance for $MgNiMo_3O_8$ is clearly two components. These features were fit using two Lorentzian curves, from which the g factor, integrated intensity, and width could be extracted. The temperature dependence of these parameters are plotted in Fig. 6 (d-f) and Fig. 7 (d-f).

We can leverage our knowledge of the stoichiometry and site occupancy in $MgNiMo_3O_8$ and the measured signals from $Ni_2Mo_3O_8$ and $MgNiMo_3O_8$ to separate the signals from the two sites. The higher intensity feature in $MgNiMo_3O_8$ corresponds to the 86 % stoichiometric octahedral fraction, while the lower intensity peak corresponds to the 14 % stoichiometric tetrahedral fraction. Subtracting the $Ni_2Mo_3O_8$ and $MgNiMo_3O_8$ signals with scaling factors for occupancy yield the single-contribution peaks (SI Fig. 2). The resonance at lower (higher) field corresponds to the tetrahedral (octahedral) component: when the scaled fraction of $Ni_2Mo_3O_8$ is subtracted from the $MgNiMo_3O_8$, the higher field feature remains.

The g-factor for the octahedral site is temperature insensitive in both $MgNiMo_3O_8$ and $Ni_2Mo_3O_8$ and remains at $\approx 2.2$ from $T = 300$ K to $T = 10$ K. In contrast, the g-factor for the tetrahedral site remains constant at $\approx 3.7$ from $T = 290$ K to $T \sim 120$ K and then increases from $T \approx 130$ K to $\approx 4.3$ as temperature decreases to $T = 10$ K.

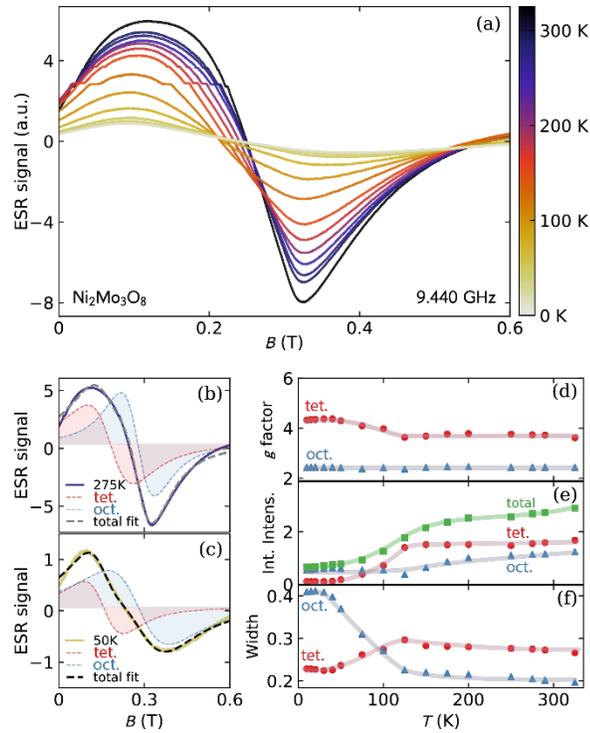

FIG. 6 (a) Temperature dependent electron spin resonance signal of $Ni_2Mo_3O_8$ in the $T = 10$ K to $T = 325$ K range. Two Lorenzian peak profiles were used to fit the data, shown for (b) $T = 275$ K and (c) $T = 50$ K, and the (d) g factor, (e) integrated intensity, and (f) width have a temperature dependence for the tetrahedral (red circles) and octahedral (blue triangles) coordination environments. Total integrated intensity is represented with green squares. Guides to the eye are drawn for panels d, e, and f.

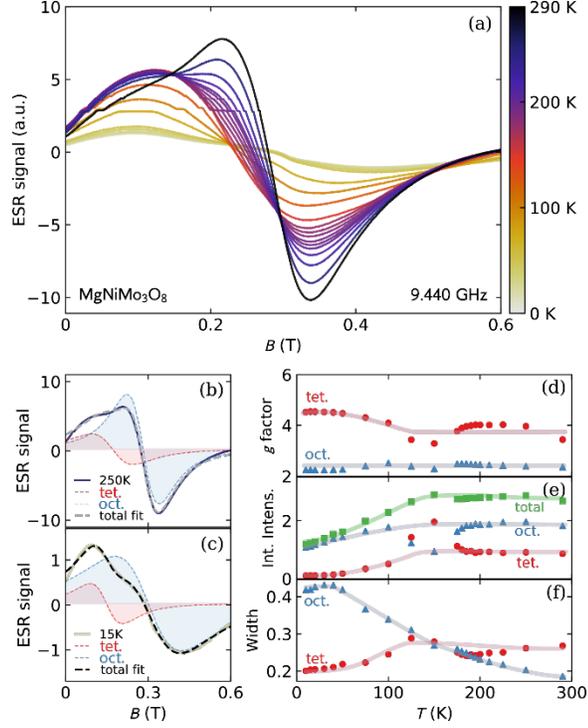

FIG. 7 (a) Temperature dependent electron spin resonance signal of $MgNiMo_3O_8$ in the $T = 10$ K to $T = 325$ K range. Two Lorenzian peak profiles were used to fit the data, shown for (b) $T = 275$ K and (c) $T = 50$ K, and the (d) g factor, (e) integrated intensity, and (f) width have a temperature dependence for the tetrahedral (red circles) and octahedral (blue triangles) coordination environments. Total integrated intensity is represented with green squares. Guides to the eye are drawn in panels d, e, and f.

Above $T = 150$ K, the octahedral data have two isosbestic points: one at 0.28 T and the other at 0.18 T. Below $T = 150$ K, there is one isosbestic point at 0.23 T. The integrated intensity for both $Ni_2Mo_3O_8$ and $MgNiMo_3O_8$ decreases from $T \sim 150$ K to $T = 10$ K.

### 8. *Single Ion Crystal Field Analysis*

Using the low temperature crystal structure, a point charge model can be used to construct the expected splitting of multielectron states for $Ni^{2+}$ on the octahedral and tetrahedral sites, Fig. 8. As expected, the trigonal distortion removes the orbital degeneracy for the tetrahedral case, but leaves the (orbitally non-degenerate) ground state of the octahedral site intact. The confluence of the trigonal crystal field with spin orbit coupling lifts the degeneracy of the ground state triplet resulting in single ion anisotropies of $\Delta = 22$ meV and $\Delta = 7.8$ meV for tetrahedral and octahedral respectively. Crucially, the low lying states on the two distinct sites are symmetry compatible, and thus can have significant exchange/superexchange interactions, in agreement with the large and negative Weiss temperature observed for $Ni_2Mo_3O_8$. Further, the single ion anisotropy of the tetrahedral site is consistent with the temperature dependent changes observed in ESR: the g-factor is expected to start to change from its high temperature to low temperature value around $0.42*\Delta = 107$ K, versus the observed $T = 110$ K. In contrast, the octahedral site would not have a local change in anisotropy until $\simeq 30$ K, a temperature at which interactions between sites are already dominant.

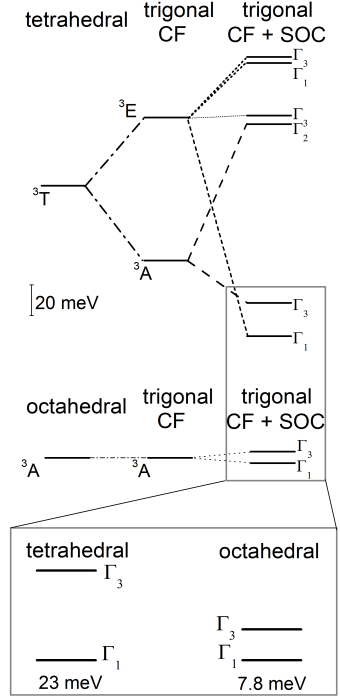

FIG. 8 Diagram of the single ion energy levels of the (left) undistorted tetrahedral and octahedral coordination environments, (middle) trigonal distortion, and (right) trigonal distortion and spin orbit coupling (SOC). Bottom: the two lowest energy states of tetrahedral and octahedral crystal field environments are similar in energy splitting and have the same $\Gamma_1$ and $\Gamma_3$ representations in $C_{3v}$, the local symmetry of both Ni ion sites.

*(a) Magnetic Structure Determination*

Magnetic Bragg peaks were identified in NPD patterns of $Ni_2Mo_3O_8$ at $T = 1.6$ K that were not present at $T = 15$ K, which is consistent with the magnetic phase transition observed in susceptibility data. These peaks were isolated by subtraction of nuclear peaks measured at the two temperatures and can be seen in Fig. 9. The largest propagation vector, $\vec{k}$, the smallest vector in real space that indexes all of the magnetic peaks is $\vec{k} = (½\ 0\ 0)$. This indicates a doubling of the unit cell in the *a* direction is necessary to describe the magnetic order. It should be noted that space group $P6_3mc$ is non-orthogonal, and this doubling is in internal *abc* directions, rather than orthogonal *xyz* directions. Representational analysis of this $\vec{k}$ vector in space group $P6_3mc$ leads to four irreducible representations: $\Gamma_1$, $\Gamma_2$, $\Gamma_3$, and $\Gamma_4$ on six basis vectors $\psi_1$ -$\psi_6$, which are summarized in Table IV. Consistent with Landau theory, only a single irreducible representation is necessary to describe the structure resulting from a second order phase transition.

TABLE IV. Irreducible representations (IR) and basis vectors (BV) for the two magnetic nickel atoms in $Ni_2Mo_3O_8$ and associated real components in the *a, b,* and *c* directions for $\vec{k} = $ (½ 0 0) in space group $P6_3mc$.

| IR | BV | atom | $m_{\parallel a}$ | $m_{\parallel b}$ | $m_{\parallel c}$ |
|---|---|---|---|---|---|
| $\Gamma_1$ | $\psi_1$ | Ni1 | 0 | -1 | 0 |
| | | Ni2 | 0 | -1 | 0 |
| $\Gamma_2$ | $\psi_2$ | Ni1 | 2 | 1 | 0 |
| | | Ni2 | 2 | 1 | 0 |
| | $\psi_3$ | Ni1 | 0 | 0 | 2 |
| | | Ni2 | 0 | 0 | -2 |
| $\Gamma_3$ | $\psi_4$ | Ni1 | 0 | -1 | 0 |
| | | Ni2 | 0 | 1 | 0 |
| $\Gamma_4$ | $\psi_5$ | Ni1 | 2 | 1 | 0 |
| | | Ni2 | -2 | -1 | 0 |
| | $\psi_6$ | Ni1 | 0 | 0 | 2 |
| | | Ni2 | 0 | 0 | 2 |

The intensity of neutrons scattering off of long range magnetic moments corresponds to the magnetic moment perpendicular to the neutron scattering vector. The tallest magnetic peak at $2\theta = 24.10°$ corresponds to the (004) reflection. The significant amount of intensity in this and related reflections means that there must be intensity in the *c* direction. There is no coefficient giving rise to intensity in the *c* direction in the $\Gamma_1$ and $\Gamma_3$ irreducible representations, so these may be discarded. Both $\Gamma_2$ and $\Gamma_4$ allow for intensity at all indexed peaks; Between the two, refinements of $\Gamma_2$ show a better fit to the data, with a statistical $\chi^2$ of 4.479, where the best fit of $\Gamma_4$ gives a $\chi^2$ of 5.502. A comparison of the statistical refinements can be seen in Table I in the SI.

With no constraints on magnitude and direction of magnetic moment, the refined magnetic structure in $\Gamma_2$ is a zig-zag structure. Three other common ordering patterns for honeycomb lattices were also explicitly tested: ferromagnetic (FM), Néel AFM, stripy AFM. In these refinements, the sign of the moment (+/-) in $c$ relative to the honeycomb lattice was constrained, but the magnitude and direction of the magnetic moment were not. The results of these refinements are shown in Fig. 9 (a-d), and the structure visualized in Fig. 9 (e, f). It is clear that (a) FM, (b) stripe AFM, and (c) Néel AFM do not fit the data as well as the zig-zag AFM structure (d-f).

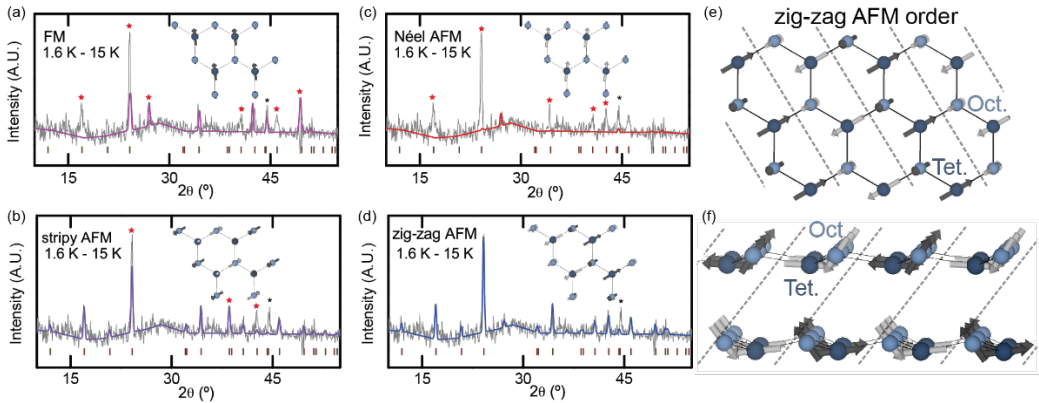

FIG. 9 Refined models with enforced (a) ferromagnetic (FM), (b) stripy antiferromagnetic (AFM), (c), Néel AFM, and (d) zig-zag AFM order on neutron powder diffraction patterns collected at $T = 1.6$ K with the nuclear contribution subtracted using measurements done at $T = 15$ K. (a) FM and (c) Néel AFM order do not have intensity at many magnetic peaks; Zig-zag AFM order results in the best fit. Red asterisks denote significant deviations of the fit from the data. The black asterisk denotes a remnant structural contribution. (e) Top-down and (f) side view of the zig-zag structure. Magnetic moment in the $+c$ ($-c$) direction are light (dark) gray, dark (light) blue atoms are tetrahedrally (octahedrally) coordinated nickel.

All combinations of larger moment on the tetrahedral site or the octahedral site, initiated with magnitude in the *c* direction or the *ab* plane, and every combination of positive and negative starting values for the coefficients of the basis vectors were refined using the nuclear-subtracted magnetic Bragg peaks with no constraints on magnitude and direction. All refinements resulted in zig-zag order. While there is no statistical difference between the $\chi^2$ metric of the quality of the refinements that have more magnitude on the octahedral or tetrahedral nickel site (the sites are indistinguishable if only the Ni atom positions are considered), it is clear from ESR data that there is a larger magnetic moment on the tetrahedral nickel.

There are two statistically identical zig-zag magnetic structures with larger magnetic moment on the tetrahedral nickel. There is strong directionality to the magnetic moment of the two sites of both. In one, an ordered moment of 1.727 $\mu_B$ on the tetrahedral site lies mainly in the *ab* plane and a moment of 1.431 $\mu_B$ on the octahedral site points primarily in the *c* direction. In the other, an ordered moment of 1.997 $\mu_B$ on the tetrahedral site points partially in the *c* direction and a moment of 0.891 $\mu_B$ on the octahedral site is mainly in the *ab* plane. The ratio of the tetrahedral to octahedral g factors (which are proportional to the magnetic moment) is 1.21 for a structure where the tetrahedral moment is primarily in the *ab* plane and 2.24 for the moment in the *c* direction. These numbers bracket the ratio of 1.8 observed in the ESR measurements at $T = 10$ K, Table V. The refinement to the structure where the tetrahedral spins lie mainly in the *ab* plane better describes the data, based on visual inspection. This solution is more intuitive, too, as one would expect the magnetic moment to be roughly the same for the two sites, as nickel is 2+ on both.

TABLE V. Values and ratios of tetrahedral to octahedral magnetic moments from ESR measured at $T = 10$ K and refinements in $\mathbf{\Gamma_2}$ to the magnetic Bragg peaks from NPD with the tetrahedral spins primarily in the $ab$ plane or the $c$ direction.

|  | $\mu_{B,Tet.}$ | $\mu_{B,Oct.}$ | $\dfrac{\mu_{B,Tet.}}{\mu_{B,Oct.}}$ |
|---|---|---|---|
| ESR T = 10 K | 4.32 | 2.43 | 1.78 |
| $ab$ plane | 1.727 | 1.431 | 1.21 |
| $c$ direction | 1.997 | 0.891 | 2.24 |

D. Discussion

The ratio of the tetrahedral site g-factor to the octahedral site g-factor determined by ESR at $T = 290$ K is 1.46, which is very close to 1.52, the ratio of the effective magnetic moments per Ni of $Ni_2Mo_3O_8$ to $MgNiMo_3O_8$ found by Curie-Weiss analysis of temperature-dependent magnetization. This further validates the agreement of the magnetic measurements and the conclusion that $MgNiMo_3O_8$ is an analog for the magnetic behavior for isolated nickels interacting on the octahedrally coordinated sublattice of the honeycomb. This ratio is also close to the ratio of the ordered magnetic moments on the tetrahedral and octahedral sites determined by NPD.

The data supports the interpretation that there is anisotropy to the magnetism on the tetrahedral site in $Ni_2Mo_3O_8$. (1) The zig-zag ordered structure shows a strong directional dependence of the magnetic moment on the two sites where the tetrahedral site has a strong $ab$ plane component, (2) the observed metamagnetism in the field-dependent magnetization (inset, Fig. 5 (a)) is a signature of anisotropy in powder samples, and has been observed in other honeycombs such as $\alpha$-$RuCl_3$ [8], and (3) the entropy recovered in heat capacity measurements is consistent with Ni on the tetrahedral site recovering $R\ln(2)$ in $Ni_2Mo_3O_8$.

The expected recovered entropy for a triangular lattice of $S = 1$ ions with three spin degrees of freedom is Rln(3) and for a honeycomb lattice (comprised of two triangular sublattices) is 2Rln(3). As summarized in Table II, $Ni_2Mo_3O_8$ recovers ≈ Rln(2) + Rln(3) and $MgNiMo_3O_8$ recovers 6.9(3) J mol$^{-1}$ K$^{-1}$ = 0.764Rln(3) of entropy. The site disorder determined by NPD places 86 % of Ni on the octahedral site in $MgNiMo_3O_8$. The theoretical change in entropy if the octahedral site were to recover Rln(2) and the tetrahedral site were to recover Rln(3) is 6.1 J mol$^{-1}$ K$^{-1}$. As this is smaller than the recovered value, it is clear that the octahedral site must be recovering Rln(3). The value of 0.76Rln(3) suggests, but does not conclusively prove, that the tetrahedral site does not recover significant entropy in $MgNiMo_3O_8$. That the entropy in $Ni_2Mo_3O_8$ recovers Rln(3) + Rln(2) strongly suggests that the tetrahedral site recovers Rln(2) of entropy, and thus has one fewer degree of freedom than the octahedral site. This implies spin anisotropy, perhaps easy-plane, which is consistent with the magnetic structure.

There are three known possible magnetic Hamiltonians which could stabilize zig-zag AFM order in $Ni_2Mo_3O_8$: (1) bond-dependent Heisenberg-Kitaev interactions [24, 48], (2) isotropic interactions where nearest neighbor (2N), next-nearest neighbor (3N), and next-next-nearest neighbor (4N) in-plane interactions are all of similar strength [12, 13], and (3) bond-dependent anisotropic interactions through ligand distortion [35].

(1) The Kitaev model requires that exchange anisotropy must be orthogonal to the Ni-Ni bond and that there are 90° interfering ligand superexchange pathways for Ising-like terms to emerge [49]. In $Ni_2Mo_3O_8$, the Ni-O-Ni bond lies along a mirror plane which precludes the necessary orthogonality. In addition, the alternating octahedral and tetrahedral coordination environments geometrically obstruct the ligand superexchange pathway.

(2) Isotropic interactions can stabilize zig-zag order when the 2N, 3N, and 4N in-plane interactions are all of similar strength. In $Ni_2Mo_3O_8$, 2N interactions are Oct.-Tet. (3.39 Å; oxygen mediated), 3N interactions are self-sublattice Oct.-Oct. and Tet.-Tet. (5.96 Å; oxygen and molybdenum mediated), and 4N are Oct. -Tet. (6.680(5) Å). $MgNiMo_3O_8$ can be viewed as a magnetically dilute analog of $Ni_2Mo_3O_8$ where the interacting magnetic atoms are predominantly structurally equivalent to the 3N interaction sublattice in $Ni_2Mo_3O_8$. While not a perfect analog, the type and relative scale of the magnetic interactions in $MgNiMo_3O_8$ is suggestive of the characteristics of the $Ni_2Mo_3O_8$ 3N interactions in the absence of the 2N interactions. The result of this magnetic dilution is a dramatic loss of interaction strength: the Weiss temperature of $MgNiMo_3O_8$ is small and positive, 6 K, indicating that the interactions are small and ferromagnetic; For comparison, the Weiss temperature of $Ni_2Mo_3O_8$ is -55 K. Thus it is likely that nearest neighbor interactions are making up the bulk of the antiferromagnetic interactions in $Ni_2Mo_3O_8$ and isotropic interactions are likely not stabilizing the zig-zag order.

(3) There are slight distortions of the octahedral and tetrahedral coordination environments from the ideal single-ion crystal field to the symmetry-adapted, spin-orbit-coupled regime. Both $Ni^{2+}$ ions are on sites with $3m$ ($C_{3v}$) symmetry, which is significantly lower point symmetry than either the $O_h$ or $T_d$ point groups in the single ion regime. As described in Fig.8, the lowest energy state in an undistorted octahedral complex is $^3A$, which decomposes into a singlet $\Gamma_1$ and doublet $\Gamma_3$ under small trigonal distortions and application of spin orbit coupling in $3m$ symmetry. The next lowest energy state is 490 meV higher. In the tetrahedral coordination, the ground state is a spin and orbital triplet, $^3T$, which decomposes into a singlet $\Gamma_1$ and doublet $\Gamma_3$ under small trigonal distortions and application of spin orbit coupling in $3m$ symmetry. It is possible that the bond-dependent interactions that occur as a result of $\Gamma_1$-$\Gamma_1$ and $\Gamma_3$-$\Gamma_3$ mixing in adjacent octahedral and tetrahedral coordination environments stabilize zig-zag order in Ni$_2$Mo$_3$O$_8$.

Bond-dependent interactions are consistent with the data collected. In particular, the rich temperature-dependent behavior in the ESR data suggest the presence of single ion anisotropy that changes with temperature: the g-factor increases between $T = 130$ K and $T = 10$ K, and below $T \approx 150$ K the amplitude of the signal decreases. This is attributable to a change the timescale of paramagnetic fluctuations to frequencies below those that ESR samples as the sample heads toward magnetic order. Additionally, the ratio of the $a$ and $c$ lattice parameters shows anisotropic changes concomitant with the temperature dependence of the ESR data.

## Conclusions

Ni$_2$Mo$_3$O$_8$ is the first realized example of an integer spin zig-zag AFM ordered honeycomb in a non-centrosymmetric space group ($P6_3mc$). Theoretical studies have predicted the existence of topological magnons in honeycomb compounds with zig-zag AFM order, and Ni$_2$Mo$_3$O$_8$ may provide an opportunity to investigate this and other topological phenomena experimentally. The zig-zag AFM order on Ni$_2$Mo$_3$O$_8$ may be stabilized by bond-dependent anisotropic exchange due to ligand distortion; the unique structure of alternating octahedral and tetrahedral Ni$^{2+}$ on the honeycomb offers fundamentally different chemistry from other nickel honeycomb compounds in existence. We have also shown that the magnetic exchanges in this material are tuneable by chemical substitution of one ion on the honeycomb, from weakly ferromagnetic (MgNiMo$_3$O$_8$) to strongly antiferromagnetic (FeNiMo$_3$O$_8$). Further studies on these materials will advance the search for realized non-trivial quantum states of matter.

ACKNOWLEDGEMENTS

This work supported by the US Department of Energy, Office of Basic Energy Sciences, Division of Materials Sciences and Engineering under Award DE-FG-02-08ER46544 to the Institute for Quantum Matter at Johns Hopkins University. Additional neutron diffraction experiments were done on the POWGEN instrument at Oak Ridge National Laboratory by Ashfia Huq. TMM acknowledges fruitful discussions with Andriy Nevidomskyy. JRM acknowledges Zachary Kelly and Christopher Pasco for useful consultations. The authors acknowledge Jiajia Wen for measuring silver on the dilution refrigerator for data analysis purposes. TMM acknowledges additional support from the Johns Hopkins University Catalyst Award. The dilution refrigerator was funded through the National Science Foundation Major Research Instrumentation Program, Grant No. DMR-0821005. Diagnostic measurements were done using instrumentation at the PARADIM bulk crystal growth facility at Johns Hopkins University, a National Science Foundation Materials Innovation Platform. Certain commercial equipment, instruments, or materials are identified in this document. Such identification does not imply recommendation or endorsement by the National Institute of Standards and Technology nor does it imply that the products identified are necessarily the best available for the purpose.

References

[1] M. Z. Hasan and C. L. Kane, *Rev. Mod. Phys.* 82, 3045 (2010).
[2] Y. Ando and L. Fu, *Ann. Rev. Condens. Matter Phys.* 6, 1 (2015).
[3] Y. Zhou, K. Kanoda, and T-K. Ng, *Rev. Mod. Phys.* 89, 025003 (2017).
[4] L. Savary and L. Balents, *Rep. Prog. Phys.* 80, 016502 (2016).
[5] B. K. Clark, D. A. Abanin, and S. L. Sondhi, *Phys. Rev. Lett.* 107, 087204 (2011).
[6] A. Kitaev, Ann. Phys. 321, 2 (2006).
[7] K. W. Plumb, J. P. Clancy, L. J. Sandilands, V. V. Shankar, Y. F. Hu, K. S. Burch, Hae-Young Kee, and Young-June Kim, *Phys. Rev. B* 90, 041112(R) (2014).
[8] J. A. Sears, M. Songvilay, K. W. Plumb, J. P. Clancy, Y. Qiu, Y. Zhao, D. Parshall, and Young-June Kim, *Phys. Rev. B* 91, 144420 (2015).
[9] H.-S. Kim, V. V. Shankar, A. Catuneanu, and H.-Y. Kee, *Phys. Rev. B* 91, 241110(R) (2015).
[10] C. A. Bridges, J.-Q. Yan, A. A. Aczel, L. Li, M. B. Stone, G. E. Granroth, M. D. Lumsden, Y. Yiu, J. Knolle, S. Bhattacharjee, D. L. Kovrizhin, R. Moessner, D. A. Tennant, D. G. Mandrus and S. E. Nagler, *Nat. Mater.* 15 (2016).
[11] A. Banerjee, C. A. Bridges, J.-Q. Yan, A. A. Aczel, L. Li, M. B. Stone, G. E. Granroth, M. D. Lumsden, Y. Yiu, J. Knolle, S. Bhattacharjee, D. L. Kovrizhin, R. Moessner, D. A. Tennant, D. G. Mandrus and S. E. Nagler, *Nat. Mater.* 15 (2016).
[12] P. H. Y. Li, R. F. Bishop, D. J. J. Farnell, and C. E. Campbell, *Phys. Rev. B* 86, 144404 (2012).
[13] A. F. Albuquerque, D. Schwandt, B. Hetényi, S. Capponi, M. Mambrini, and A. M. Läuchli, *Phys. Rev. B* 84, 024406 (2011).
[14] Y-Z You, I. Kimchi, and A. Vishwanath, *Phys. Rev. B* 86, 085145 (2012).

[15] J. Reuther, R. Thomale, and S. Rachel, *Phys. Rev. B* 90, 100405(R) (2014).
[16] A. Biffin, R. D. Johnson, I. Kimchi, R. Morris, A. Bombardi, J. G. Analytis, A. Vishwanath, and R. Coldea, *Phys. Rev. Lett.* 113, 197201 (2014).
[17] T. Takayama, A. Kato, R. Dinnebier, J. Nuss, H. Kono, L. S. I. Veiga, G. Fabbris, D. Haskel, and H. Takagi, *Phys. Rev. Lett.* 114, 077202 (2015).
[18] S. C. Williams, R. D. Johnson, F. Freund, S. Choi, A. Jesche, I. Kimchi, S. Manni, A. Bombardi, P. Manuel, P. Gegenwart, and R. Coldea. *Phys. Rev. B* 93, 195158 (2016).
[19] H-C. Jiang, Z-C. Gu, X-L. Qi, and S. Trebst, *Phys. Rev. B* 83, 245104 (2011).
[20] Y. Singh, S. Manni, J. Reuther, T. Berlijn, R. Thomale, W. Ku, S. Trebst, and P. Gegenwart, *Phys. Rev. Lett.* 108, 127203 (2012).
[21] F. Ye, S. Chi, H. Cao, B. C. Chakoumakos, J. A. Fernandez-Baca, R. Custelcean, T. F. Qi, O. B. Korneta, and G. Cao, *Phys. Rev. B* 85, 180403(R) (2012).
[22] J. Chaloupka, G. Jackeli, and G. Khaliullin, *Phys. Rev. Lett.* 110, 097204 (2013).
[23] V. M Katukuri, S. Nishimoto, V. Yushankhai, A. Stoyanova, H. Kandpal, S. Choi, R. Coldea, I. Rousochatzakis, L. Hozoi and J. van den Brink, *New J. Phys.* 16 (2014).
[24] K. A. Modic, T. E. Smidt, I. Kimchi, N. P. Breznay, A. Biffin, S. Choi, R. D. Johnson, R. Coldea, P. Watkins-Curry, G. T. McCandless, J. Y. Chan, F. Gandara, Z. Islam, A. Vishwanath, A. Shekhter, R. D. McDonald and J. G. Analytis, *Nat. Comm.* 5 (2014).
[25] J. Chaloupka and G. Khaliullin, *Phys. Rev. B* 94, 064435 (2016).
[26] A. F. Albuquerque, D. Schwandt, B. Hetényi, S. Capponi, M. Mambrini, and A. M. Läuchli, *Phys. Rev. B* 84, 024406 (2011).
[27] F. Mezzacapo and M. Boninsegni, *Phys. Rev. B* 85, 060402(R) (2012).
[28] J. G. Rau, E. K.-H. Lee, and H.-Y. Kee, *Phys. Rev. Lett.* 112, 077204 (2014).
[29] Z. Nussinov and J. van den Brink, *Rev. Mod. Phys.* 87, 1 (2015).
[30] E. A. Zvereva, M. I. Stratan, A. V. Ushakov, V. B. Nalbandyan, I. L. Shukaev, A. V. Silhanek, M. Abdel-Hafiez, S. V. Streltsovbh and A. N. Vasilievahi, *Dalton Trans.* 45, 17 (2016).
[31] H. Liu and G. Khaliullin, *Phys. Rev. B* 97, 014407 (2018).
[32] S. A. Owerre, *J. Phys. Condens. Matter* 28, 38 (2016).
[33] H. Lee, S. B. Chung, K. Park, J.-G. Park, *arXiv:1712.09801* (2017).
[34] S.P. McAlister and P. Strobel, *J. Magn. Magn. Mater.* 20, 3 (1983).
[35] E. A. Zvereva, M. I. Stratan, Y. A. Ovchenkov, V. B. Nalbandyan, J.-Y. Lin, E. L. Vavilova, M. F. Iakovleva, M. Abdel-Hafiez, A. V. Silhanek, X.-J. Chen, A. Stroppa, S. Picozzi, H. O. Jeschke, R. Valenti, and A. N. Vasiliev, *Phs. Rev. B* 92, 144401 (2015).
[36] E. M. Seibel, J. H. Roudebush, H. Wu, Q. Huang, M. N. Ali, H. Ji and R. J. Cava, *Inorg. Chem.* 52, 23 (2013).
[37] R. Berthelot, W. Schmidt, S. Muir, J. Eilertsen, L. Etienne, A. W. Sleight and M. A. Subramanian, *Inorg. Chem.* 51, 9 (2012).
[38] A.C. Larson and R.B. Von Dreele, *Los Alamos National Laboratory Report, LAUR* 86-748 (2000).
[39] B.H. Toby, *J. Appl. Cryst.* 34, 210 (2001).
[40] Rodrigues-Carvajal, *Phys. B* 192, 55 (1993).
[41] Wills, A. S. *Phys. B*, 276 (2000).
[42] K. Momma and F. Izumi, *J. Appl. Crystallogr.* 44 (2011).
[43] A. Tari, *The Specific Heat of Matter at Low Temperatures* (Imperial College Press, London, 2003).

[44] T. Kurumaji, Y. Takahashi, J. Fujioka, R. Masuda, H. Shishikura, S. Ishiwata, and Y. Tokura, *Phys. Rev. B* 95, 020405(R) (2017).
[45] M. Hutchings, *Solid State Physics*, 16 , 227 (1964).
[46] A. Scheie, PyCrystalField, (2018), GitHub repository, https://github.com/asche1/PyCrystalField
[47] A. Abragam and B. Bleaney, *Electron Paramagnetic Resonance of Transition Ions*, 1st ed. (Clarendon Press, Oxford, 1970).
[48] H. B. Cao, A. Banerjee, J.-Q. Yan, C. A. Bridges, M. D. Lumsden, D. G. Mandrus, D. A. Tennant, B. C. Chakoumakos, and S. E. Nagler, *Phys. Rev. B* 93, 134423 (2016).
[49] G. Jackeli and G. Khaliullin, *Phys. Rev. Lett.* 102, 017205 (2009).

SUPPLEMENTARY INFORMATION

The fit statistic $\chi^2$ is calculated as in Eq. (1) where N is the number of points less the number of refined parameters (for all fits, N >> number of refined parameters), $I_{C,i}$ is the calculated intensity at each point *i*, $I_{O,i}$ is the observed intensity at each point *i*, and $\sigma$ is the standard deviation.

$$\chi^2 = \frac{1}{N}\frac{\Sigma_i(I_{C,i}-I_{O,i})^2}{\sigma^2[I_{O,i}]} \tag{1}$$

The fit statistic *wRp* is calculated as in Eq. (2) where the weighting factor $w_i = 1/\sigma^2[I_{O,i}]$

$$wRp = \sqrt{\frac{\Sigma_i w_i(I_{C,i}-I_{O,i})^2}{\Sigma_i w_i(I_{O,i})^2}} \tag{2}$$

The fit statistic **Rp** is calculated as in Eq. (3)

$$Rp = \sqrt{\frac{N}{\Sigma_i w_i(I_{O,i})^2}} \tag{3}$$

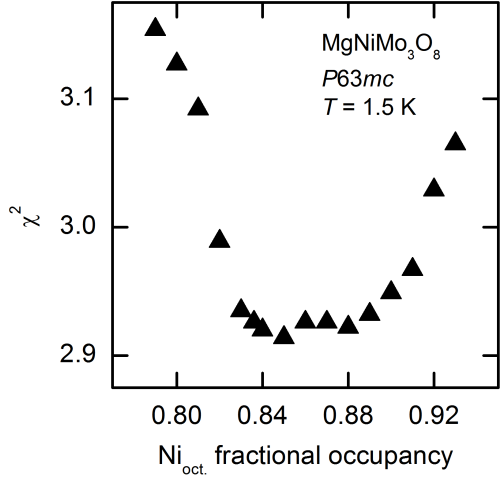

FIG. 10 Dependence of the fit statistic $\chi^2$ on the fractional occupancy of nickel on the octahedral site of the Mg-Ni honeycomb lattice. Total occupancy of the site was held at 1.

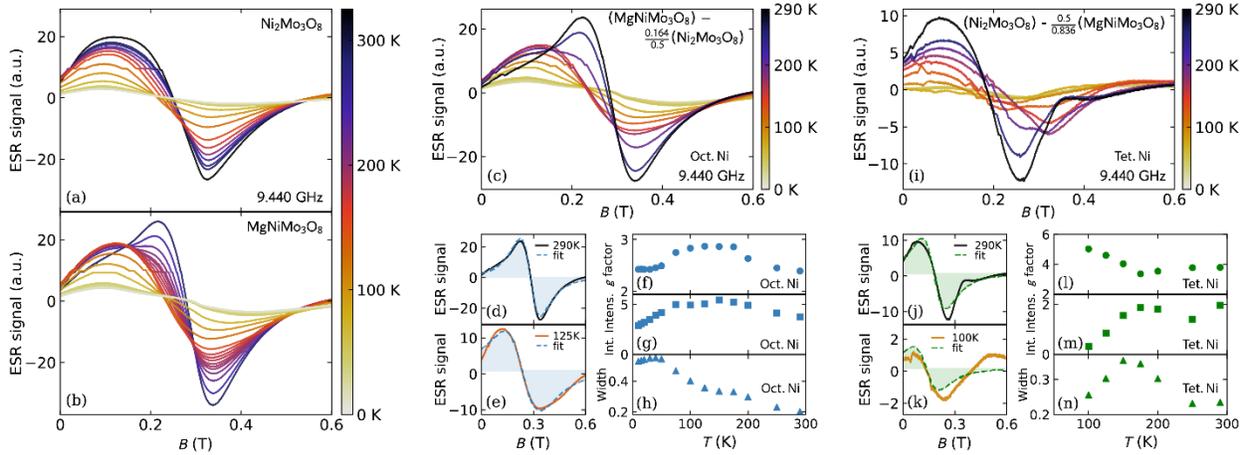

FIG. 11 (a) Temperature dependent electron spin resonance signal of (a) $Ni_2Mo_3O_8$ in the $T = 10$ to $T = 325$ K range, and (b) $MgNiMo_3O_8$ in the $T = 10$ to $T = 290$ K range measured at a frequency of 9.440 GHz. (c)(i) Plot of the octahedral (tetrahedral) component of the $MgNiMo_3O_8$ ($Ni_2Mo_3O_8$) data, and fits of a Lorentzian profile to data at (d)(j) 290 K and (e)(k) 125 K (100 K). Plots of (f)(l) g-factor, (g)(m) integrated intensity, and (h)(n) width parameters of fits at all measured temperatures.

TABLE VI Refinement statistics for fits using the irreducible representations $\Gamma_2$ and $\Gamma_4$ on the magnetic peaks in neutron powder diffraction patterns of $Ni_2Mo_3O_8$. Initialization

of refinements with more magnitude in the *c* direction or the *ab* plane resulted in subtly different solutions.

|  | $\Gamma_2$ | | $\Gamma_4$ | |
| --- | --- | --- | --- | --- |
|  | *c* direction | *ab* plane | *c* dir. | *ab* plane |
| $\chi^2$ | 4.479 | 4.479 | 5.546 | 5.502 |

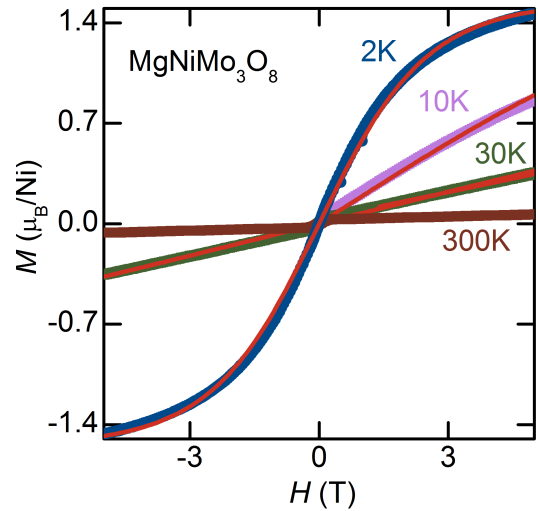

FIG. 12 Field-dependent magnetization of MgNiMo$_3$O$_8$ measured at 2 K, 10K, 30K, and 300 K. Red curves represent fits of a Brillouin function (Eq. 5) to the data. Fit values are summarized in Table II.

TABLE VII Refined values and fit statistics for fits of a Brillouin function to field-dependent magnetization of MgNiMo$_3$O$_8$ at $T = 2$ K, 10K, and 300 K.

| $T$ (K) | J | $R^2$ |
| --- | --- | --- |
| 2 | 0.7751(9) | 0.99874 |
| 10 | 1.051(18) | 0.99504 |
| 30 | 1.157(2) | 0.9942 |

Magnetization is defined as:

$$M = g_j \times J \times B_j \tag{4}$$

Where the Brillouin function $B_J$ as a function of angular momentum *J* is:

$$B_j = \frac{2J+1}{2J} coth\left(\frac{2J+1}{2J} \times t\right) - \frac{1}{2J} coth\left(\frac{t}{2J}\right) \tag{5}$$

And the ratio of the magnetic and thermal energies is:

$$t = \frac{f \times g_j \times J}{kT} \times H \quad (6)$$

Where $M$ is magnetization, $H$ is applied field, and $g_j$ is held to the spin-only value of 2.

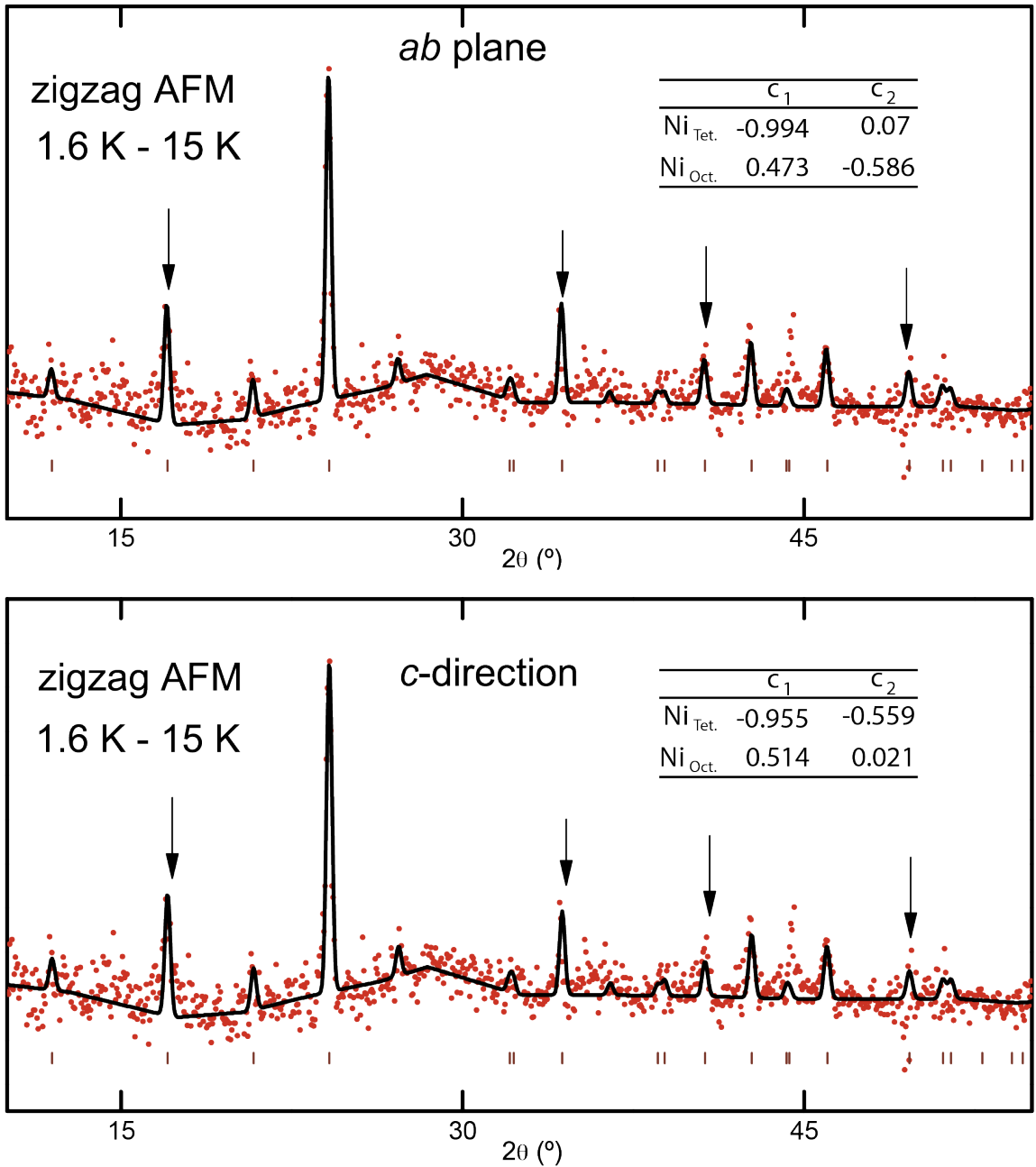

FIG. 13 Refinements to the magnetic contribution to NPD patterns. Top panel: tetrahedral magnetic moments in the $ab$ plane. Bottom panel: tetrahedral magnetic moment in the $c$

direction. Inset tables show the refined coefficients for the basis vectors for each refinement. Arrows identify peaks with significant differences between the two refinements. Visualizations of these structures can be seen in Fig. 5. The coefficient $c_1$ operates on a basis vector in the *ab* plane, $c_2$ on a basis vector in the *c* direction.

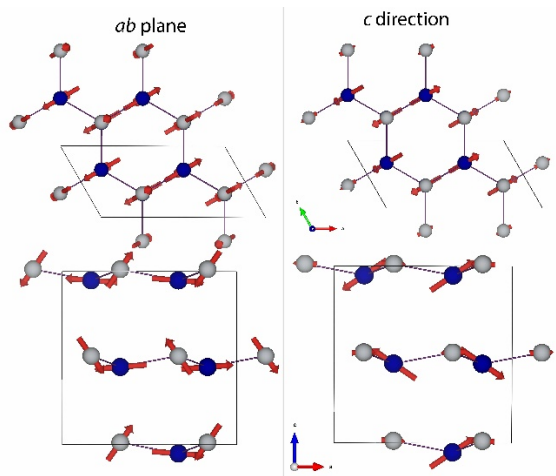

FIG. 14 Visualization of magnetic structures shown in Fig. 4. Left panel: tetrahedral magnetic moment is in the ab plane, right panel: tetrahedral magnetic moment in the c direction.

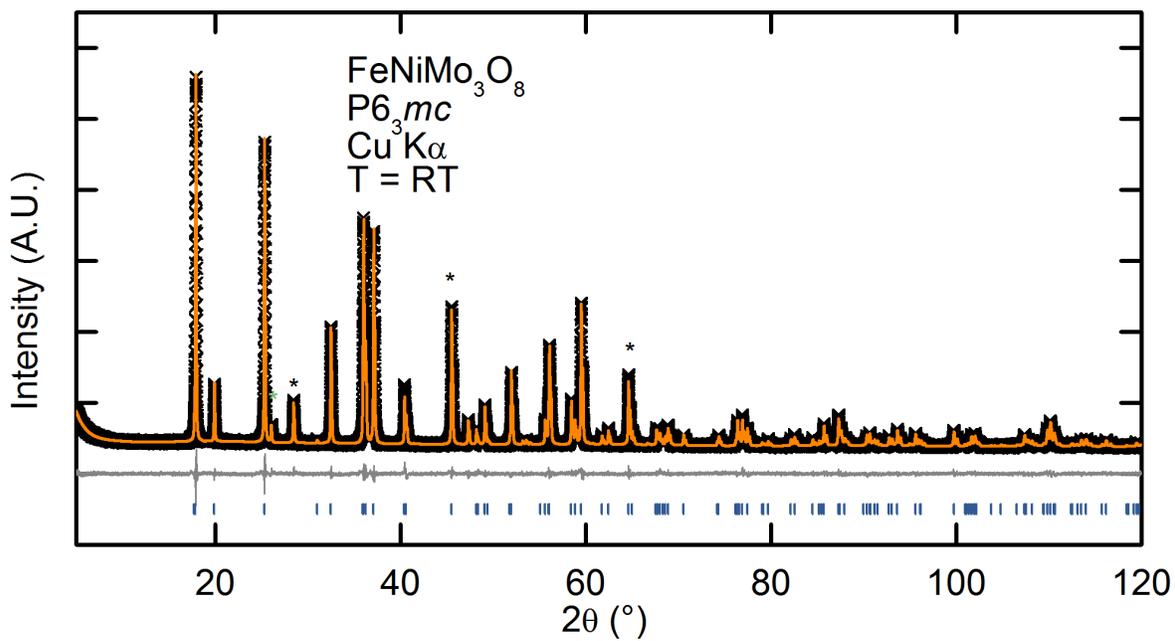

FIG. 15 Rietveld refinement of $P6_3mc$ to a room temperature PXRD pattern collected on $FeNiMo_3O_8$, measured with Cu K$\alpha$ radiation. Black asterisks denote a Si standard, and a green asterisk denotes a 1.6 % $MoO_2$ impurity.

Additional Information on PyCrystalField:

The crystal electric field (CEF) Hamiltonian can be written as $H_{CEF} = \sum_{nm} B_n^m O_n^m$ where $O_n^m$ are Stevens Operators and $B_n^m$ are multiplicative factors called CEF parameters. To calculate the energy level splittings, we computed the single-ion eigenstates using PyCrystalField, which can be downloaded at https://github.com/asche1/PyCrystalField[1]. This code, based on Hutchings (1964) [2], estimates the CEF Hamiltonian by treating ligands as point charges using the Stevens Operators formalism. To fully account for spin-orbit interactions, we calculated the single-ion Hamiltonian in the intermediate coupling scheme by expressing the crystal fields as interacting the orbital angular momentum $L$, and adding spin orbit coupling (SOC) $H_{SOC} = \lambda S \cdot L$ non-perturbatively to the Hamiltonian so that $H = H_{SOC} + H_{CEF}$.

From here, the eigenvalues and eigenvectors are calculated by diagonalizing the Hamiltonian. For $Ni^{2+}$, Hund's rules dictate that $S$=1 and $L$=3, so the Hamiltonian is written as a 21x21 matrix. Diagonalizing the matrix gives the energy splitting of the multiplets, which are shown in Fig. 8. Values of $\lambda$ and $Ni^{2+}$ radial integrals were taken from [3].

References

[1] A. Scheie, PyCrystalField, (2018), GitHub repository, https://github.com/asche1/PyCrystalField
[2] M. Hutchings, *Solid State Physics,* 16 , 227 (1964).
[3] A. Abragam and B. Bleaney, *Electron Paramagnetic Resonance of Transition Ions*, 1st ed. (Clarendon Press, Oxford, 1970).